\documentclass[titlepage,12pt]{article}
\usepackage{amssymb,epsfig,pslatex}
\usepackage{color}
\usepackage{cite}
\usepackage{colortbl}
\def\dsp#1{\displaystyle{#1}}
\def\ttbar{{t \bar t}}

\newcommand{\rmi}{\rm i}
\newcommand{\be}{\begin{equation}}
\newcommand{\ee}{\end{equation}}
\newcommand{\beq}{\begin{equation}}
\newcommand{\eeq}{\end{equation}}
\newcommand{\bea}{\begin{eqnarray}}
\newcommand{\eea}{\end{eqnarray}}
\renewcommand{\thefootnote}{\small\fnsymbol{footnote}}
\begin{document}
\begin{titlepage}
  \begin{flushright}
    FTUV-11-2705
  \end{flushright}
\vspace{0.01cm}
\begin{center}
{\LARGE {\bf Top quark tensor couplings}}  \\
\vspace{2cm}
{\large{\bf Gabriel A. Gonz\'alez-Sprinberg$^1$\footnote{Email:
{\tt gabrielg@fisica.edu.uy}},
 Roberto Martinez$^2$\footnote{Email:
{\tt remartinezm@unal.edu.co}},
 Jorge Vidal$^3$\footnote{Email: {\tt vidal@uv.es}}
}}
\par\vspace{.5cm}
$^1$Instituto de F\'\i sica, Facultad de Ciencias, Universidad de la
Rep\'ublica, Igu\'a 4225, Montevideo 11600, Uruguay.\\
\par\vspace{.5cm}
$^2$ Departamento de F\'\i sica,  Universidad Nacional de Colombia,
Bogot\'a, Colombia.\\
\par\vspace{.5cm}
$^3$ Departament de F\'\i sica Te\`orica Universitat de Val\`encia, E-46100
Burjassot,Val\`encia, Spain
and
IFIC, Centre Mixt Universitat de Val\`encia-CSIC, Val\`encia, Spain.\\
\par\vspace{1cm}
{\bf Abstract}\\
\parbox[t]{\textwidth}
{
We compute the real and imaginary parts of the one-loop electroweak
contributions to the left and right tensorial anomalous couplings of
the $tbW$
vertex in the Standard Model (SM). For both  tensorial couplings we find
that the real part of the  electroweak SM correction is close to 10$\%$ of
the leading contribution given by the QCD gluon exchange. We also find that
the  electroweak real and imaginary parts for the anomalous right coupling
are almost of the same order of magnitude. The one loop SM prediction for the real
part of the left coupling is close to the 3$\sigma$ discovery limit derived
from  $b\rightarrow s \gamma$. Besides, taking into account that the
predictions of new physics interactions are also at the  level of a few
percents when compared with the one loop QCD gluon exchange,  these
electroweak corrections  should be taken into account in order to
disentangle new physics effects from the standard ones.
These anomalous tensorial couplings of the top quark will be investigated
at the LHC in the near future where sensitivity to these contributions may
be achieved.
}
\end{center}
\vfill
\end{titlepage}
%
%
\setcounter{footnote}{0}
\renewcommand{\thefootnote}{\arabic{footnote}}
\setcounter{page}{1}
\section{Introduction}
\label{Intro} \quad
Top quark physics at the Large Hadron Collider (LHC) is an important
scenario for  testing physics above the electroweak scale.  No deviation
from the predictions of the Standard Model (SM) is found in top data
\cite{Varnes:2008tc}, nowadays dominated by the Tevatron experiments.  This
situation may change with the LHC already running and taking data. The SM
dominant decay mode $t\rightarrow b W^+$ will be precisely measured at the
LHC and sensitivity beyond tree level SM will be achieved.
It is generally believed that, due to its large mass, physics of the top
quark will be useful to probe new theories above the electroweak scale
\cite{new,Bernreuther:2008ju}.
At the LHC, top new physics may show up in new top quark decay channels
or in the measurement of the top standard and anomalous
couplings\cite{Bernreuther:2008ju, gerber}. In renormalizable theories, the
anomalous couplings appear as quantum corrections, as it is the case for the
SM, and also in many new physics theories. In a model independent approach
there are two ways of parameterize the unknown physics at high scales. One
is the  effective Lagrangian method\cite{buch} which is a way to describe
low energy physics effects originated at a higher energy scale. These
effects are parameterized with non-renormalized terms invariant under the SM
gauge symmetry $SU(3)_c \times SU(2)_L \times U(1)_Y$  and written in terms
of the low energy (standard) particle spectrum fields. It is assumed that
the new particles spectrum lies at an energy scale well above the
electroweak scale. 
The other way is just by assuming the most general form of the Lorentz
structure for the $tbW$ amplitude.  There are many terms in the effective
Lagrangian that may give contributions to the same Lorentz structure in the
vertex, in particular to the tensorial couplings we are interested in.
Besides,  some of those terms can be rewritten by using the equations
of motion, so the identification of the effective Lagrangian terms with the form
factors is not direct nor unique. In this paper we will use the second
approach, parameterizing in the most general way the $tbW$ amplitude.
This approach has the advantage, over the effective Lagrangian approach,
that it does not break down even if any relatively light particles, as new
scalars, for example, come into the game.

Some effects related to the top anomalous couplings, both in the $t
\rightarrow b W^+$ polarized branching fractions --for the three helicity W
possibilities-- and in single top production at the Tevatron and at the LHC, have already
been studied in the recent years
\cite{Bernreuther:2008ju,AguilarSaavedra:2008gt}.
However, at the LHC it will be possible to have new suitable observables in
order to perform precise measurements of  the anomalous couplings. Some
aspect of this top quark physics have also been investigated in models with
an extended Higgs sector, technicolor models, supersymmetry models and
Little Higgs models \cite{bernr}.

The anomalous couplings are gauge invariant quantities so one can think of
testing the SM predictions and new theories  through observables that are
directly sensitive to them. In fact, not only top branching fractions and
cross sections are predictions of the models to be confronted with data but,
in the same way as in the past  the anomalous magnetic moment for the
electron gave the first success of quantum field theory,  also these  $tbW$
gauge invariant tensorial couplings can be used to check the predictions of
new physics theories. One loop QCD and electroweak contributions to the top
branching fractions for polarized W's have been studied in the frame of the
SM \cite{do}. These contributions, and the corresponding measurements, have
no special sensitivity to the anomalous couplings which enter in the
observables as small corrections. The explicit dependence of the polarized
branching fractions on the  anomalous couplings have been computed in
\cite{chen,AguilarSaavedra:2006fy}, where also the sensitivity  to them has
been considered. Specific observables that are directly proportional to the
tensorial couplings have been studied in \cite{AguilarSaavedra:2006fy, ag,
AguilarSaavedra:2007rs, AguilarSaavedra:2008gt} and more recently new
observables were presented in \cite{pepe-ag}.

In this paper we compute the electroweak SM contribution to the left and
right ``magnetic'' tensorial couplings of the {\it tbW} vertex and we discuss their observable
effects; we also compare this contributions with some new physics
predictions considered in the literature. These CP-conserving pieces of the
$tbW$ vertex are different from zero only at one loop in the SM and the same
is true in many extended models. The QCD gluon-exchange contribution to the
tensorial couplings is the dominant one and has been reported in the
literature only for the right coupling \cite{Jezabek:Czarnecki,li}. The left tensorial coupling is proportional to the
bottom quark-mass due to the chirality flipping  property of this coupling  and by the fact that it only couples to a right b-quark. For these reasons it is suppressed and  it is generally assumed to be negligible. However,  the measurement of both of them  appears as  feasible in dedicated observables computed for top production at the LHC.
  
For the right tensorial  $tbW$ coupling the comparison with the SM is
usually performed  in the literature by taking as a reference only the one
loop QCD contribution. The most promising new physics  models predict a few
percents deviation from this QCD-value. However, as it is shown in this
paper, we found that the  electroweak contribution is also at the level of
$10\%$  with respect to the leading gluon exchange,  and should be taken
into account when comparing with data.  Detailed studies will be necessary
in order to disentangle new physics contributions from the electroweak
standard ones.
 
In  section 2 we define the anomalous couplings and we review  the
theoretical as well as the experimental status for the physics for which they are involved. In section
3 our computation is presented and in the final section we present our
conclusions.

\section{The tensorial $tbW$ vertex: experiment, SM and beyond}
\label{secaom} \quad
For on-shell particles, the most general amplitude ${\cal M}_{tbW}$ for the
decay $t(p) \to b(k) \, W^{+}(q)$
can be written in the following way:
\be
{\cal M}_{tbW^+} = - \frac{e}{\sin\theta_W\sqrt 2}\;
\epsilon^{\mu *} \, {\bar u}_b \left[\gamma_\mu\, \left(V_L \,P_L +
V_R \, P_R\right) + \frac{{\rmi} \sigma_{\mu\nu}  q^\nu}{m_W}
\left(g_L\, P_L  + g_R\, P_R\right)
\right] u_t \, ,
\label{eqq:1}
\ee
with $P_{L,R}=(1\mp\gamma_5)/2$; $p$, $k$ and
$q=p-k$  denote  the
top, bottom   and   $W$ boson four momenta, respectively. The tensorial
left and right magnetic moments are $g_L$ and $g_R$ respectively. The
tree level SM couplings are $V_R = 0$, $V_L = V_{tb} $ (the
Cabibbo-Kobayashi-Maskawa, CKM, matrix element),  $g_R = 0$ and $g_L =0$.
This expression for the amplitude, written in terms of the most general form
factors, is appropriate for  a model independent analysis of the tbW
amplitude.   The anomalous form factors
$g_R$ and $g_L$ are chirality flipping and   dimensionless functions of
$q^2$. For all particles on-shell, as can be assumed for the top decay, we
have  $q^2 = M_W^2$.  Besides, these dipole moments are gauge independent
quantities and may be measured with appropriate observables.

These  form factors are generated by quantum corrections in the SM. In
renormalizable theories, such as some extensions of the SM, $V_R$ can appear
at tree-level while tensor couplings $g_R$ and $g_L$, are induced as one loop quantum corrections. 
Values of $|V_L|$ different from the ones given by the global fit \cite{pdg} in the SM  $V_{tb} \simeq 1$, 
that we assume, are not experimentally excluded \cite{Alwall:2006bx} and they are still an open
window to test new physics. This issue (and also possible deviations of
$V_R = 0$) will not be the object of our work, where we concentrate only on $g_R$
and $g_L$.

In renormalizable theories, the tensorial couplings are  finite quantum
corrections quantities that do not receive contributions from
renormalization counter-terms at one loop. In addition, contrary to what
happens for the $V_R$ form factor, the tensors $g_{L,R}$ couplings are
infrared safe quantities. One loop QCD corrections generate  the leading
contribution to the tensorial couplings $g_R$ and $g_L$. This one loop QCD
gluon exchange contribution to $g_R$  was computed in \cite{li} and they
found the value $g^{QCD}_R = - 6.61\times 10^{-3}$. Direct observables with sensitivity
to $g_R$ will be accessible to the
LHC experiments as was discussed in \cite{ag}. The left tensorial
coupling term  couples a right b-quark and thus it is proportional to $m_b$.
For this reason, it is generally believed that the SM value for $g_L$ is much smaller than
the one for $g_R$. We will show that this in not exactly the case.

New physics signals can also show up in the analysis of the top decay $t\to
bW^+$. In particular,  significant deviations from the  SM predictions for
$g_R$ and $g_L$ may be found. However, the SM values are only known for $g_R$
up to one loop in QCD while the prediction for $g_L$ is not published.
Moreover, most of the analysis frequently assume real values for  $g_R$ and $g_L$.
The quantum corrections coming from the SM should be under known in
order to discriminate the SM and new physics contributions from data. 

Let us briefly review  the  experimental status for the constraints on
these tensor couplings.  Indirect limits on $g_L$ and $g_R$  can be
obtained from $b\rightarrow s \gamma$ in the measured branching ratio
$B({\bar B}\to X_s \gamma)$. The results from a recent analysis
\cite{Grzadkowski:2008mf} are  given in the first line of Table 1.

\begin{table}[hbt]{\centering
\caption{Bounds on $g_R$ and $g_L$. The first line shows the indirect
limits from $b\rightarrow s \gamma$. The second and third lines are limits
obtained from
simulations for the LHC. The last two lines show 3 $\sigma$ discovery limits
intervals: fourth line limits are from simulations for the LHC and  the
last one is from $b\rightarrow s \gamma$.
}
\begin{center}
\begin{tabular}{||c|c|c|c||}
\hline\hline
Reference& &$g_R$ bound & $g_L$ bound\\ \hline
\cite{Grzadkowski:2008mf} & 95\%C.L.&$-0.15 < g_R < 0.57$&$ -0.0015 < g_L
< 0.0004$\\ \hline
\cite{AguilarSaavedra:2007rs}&$2 \sigma$ & $ -0.026 \leq g_R \leq  0.031 $&
$-0.058 \leq g_L \leq 0.026$\\ \hline
\cite{AguilarSaavedra:2008gt} &$1 \sigma$& $-0.012 \leq g_R \leq 0.024 $&$
-0.16 \leq g_L \leq 0.16$ \\ \hline\hline
\multicolumn{2}{||c|}{}&$g_R$ discovery limit&$g_L$ discovery limit\\ \hline
&&&$Re(g_L)\geq0.051$ or \\[-7pt]
\cite{pepe-ag} &$3 \sigma$&\raisebox{2ex}{$|Re(g_R)|\geq 0.056$}& $Re(g_L)\leq -0.083$\\ \cline{3-4}
&& $ | Im ( g_R ) |   \geq 0.115$ & $  | Im ( g_L ) | \geq 0.065 $\\ \hline
&&$Re(g_R)\geq 0.76$ & $Re(g_L)\geq 0.0009$ or\\
\cite{pepe-ag,Grzadkowski:2008mf} & $3 \sigma $ & or & $ Re (g_L)\leq -0.0019$ \\ \cline{4-4}
&& $ Re(g_R)\leq -0.33$ & $  | Im ( g_L ) | \geq 0.006$\\ 
\hline\hline
\end{tabular}
\end{center}}
\label{bounds}
\end{table}

The constraints on $g_L$ are much stronger than those on $g_R$ due to the
chiral  $m_t/m_b$ enhancement factor 
which comes together with the $g_L$ coupling in the $B$-meson decay
amplitude. These bounds  are obtained assuming that all anomalous couplings
are real and that only one of them is non zero at a time.

The top width $\Gamma_t$ is an observable which is sensitive to the absolute strength of the $tbW$ vertex but with no particular sensitivity to the anomalous couplings. Another test of the Lorentz structure of the $tbW$ amplitude is the measurement of the polarized decay fractions $Br(t\to b W^+_\lambda)$ into $W^+$ bosons of helicity $\lambda = 0,\mp 1$.The SM values for this W-polarized widths are known up to one loop QCD and electroweak corrections \cite{do} and the contribution to  the helicity fractions from the anomalous couplings defined in Eq.[\ref{eqq:1}] were computed in \cite{chen,AguilarSaavedra:2006fy}. However, these fractions are sensitive only to ratios of the couplings. Other observables can be obtained from the single top production at the  LHC \cite{AguilarSaavedra:2008gt}. From a combined analysis of the single top cross section and the three helicity fractions, the four anomalous couplings in Eq.[\ref{eqq:1}] can be determined. Also, $g_R$ could be measured from the energy and angular distributions for polarized semi-leptonic and hadronic top-quark decays as was studied in \cite{Tsuno:2005qb}.

The Tevatron found no deviation from the SM in top quark physics. It has also investigated the anomalous couplings based on the single top quark production cross section \cite{Abazov:Abazov}. 
These are the first
direct experimental bounds but they are not competitive with the indirect
ones already mentioned.
In the near future  the LHC high statistics data on top quark
decays  will allow the direct determination of
the tensor couplings  $g_L$ and $g_R$ within a few percent level.
In particular,  in simulations for
$\ttbar$ production and decay in dileptonic \cite{Hubaut:2005er} and
lepton plus jets channels
\cite{AguilarSaavedra:2007rs,Tsuno:2005qb,Hubaut:2005er},   forward-backward
asymmetries
\cite{AguilarSaavedra:2007rs} and a double angular distribution in $t$-quark
decay \cite{Tsuno:2005qb} were studied.
In \cite{AguilarSaavedra:2007rs}, with only one non standard coupling
different from zero at a time,
intervals for detection or exclusion at two standard
deviations (both statistical and systematic uncertainties included) for
$g_L$ and $g_R$ were predicted for the future LHC data; they are
shown in the second line of Table 1.
The LHC will possibly improve the
sensitivity to $g_R$ by an order of magnitude
when compared to the indirect bounds from $b\rightarrow s \gamma$.
A combined fit, using the four couplings $V_L$, $V_R$, $g_R$ and $g_L$ as
parameters, and  taking into account
the expected uncertainties at the LHC for top-quark decay
was presented in  \cite{AguilarSaavedra:2008gt} and it is shown in the third line of Table 1.
The sensitivity to $g_R$ shown in the second and third lines of Table 1 is similar to the results of \cite{Tsuno:2005qb,Hubaut:2005er} where $\ttbar$ production and decay into leptonic and dileptonic decay channels were analyzed. 

Recently, new helicity fractions of the W where defined and investigated
for polarized top decays.
The spin matrix for polarized top decays was obtained in \cite{pepe-ag};
they also  considered
new observables derived from the  normal and  transverse $W^+$ polarization
fractions.
A similar approach in asymmetry  observables was also studied in Tau
physics in recent years  \cite{nos:nos1}. The three different sets of W
helicity fractions  defined for the
polarized top quark were shown to open the possibility of new observables
particularly sensitives to both the real and imaginary parts of the tensor
couplings.
They compute the $3 \sigma$ discovery limits for $g_R$ and $g_L$
assuming that they are either real or purely imaginary and allowing only one coupling to be different from zero at a time. The exclusion intervals are
shown in Table 1 in the fourth line.
As a reference for the comparison of the potential of the LHC they also
derived the 3$\sigma$ discovery limits from $b\rightarrow s \gamma$ in \cite{Grzadkowski:2008mf}; this is shown in the last line of Table 1.

It is generally believed that beyond the SM theories will be probed in top
quark physics.
These theories, in general, not only will induce non zero values of $g_R$
and $g_L$ but also may be responsible for new exotic decay modes. The top
dominant decay mode, $t\rightarrow b W^+$, was investigated in many
extension of the SM. In particular, the decay rate and polarized decay
fractions were studied in many theories such as the Two Higgs Doublet Model
(2HDM),  the minimal supersymmetric SM (MSMM) and top-color assisted
technicolor (TC2). These results were reviewed in  \cite{bernr} where some of
the anomalous couplings were computed for these models.
For the first two of them they found the general feature that $|g_R| \gg
|g_L|$ and $|Re \;(g_R)| \gg |Im\; (g_L)|$. Besides, they found that values of $g_R$ up to $0.5\times 10^{-3}$  are possible for low $\tan\beta$, while only $0.2\times 10^{-3}$ is expected for higher values of  $\tan \beta$. Notice that these two last figures represent $8\%$ and $3\%$, respectively, of the leading one loop gluon contribution. For TC2 models they showed that values for $g_R$ as big as $0.01$ can
be expected, and this represents $150\%$ of the one loop gluon contribution. The general feature  $|g_R| \gg |g_L|$ and $|Re ( g_R ) |\gg |Im ( g_L ) |$ is also true for the SM, as it will be shown in the following sections.

\section{Electroweak corrections to the anomalous couplings $g_R$ and
$g_L$}

In the SM, at one loop, there is only one topology for the vertex correction diagrams that contribute to the anomalous $g_{R}$ and $g_L$
couplings. This is shown in Figure 1(a) and we will denote this
diagram as ABC
using the name of the particles circulating in the loop. The QCD one
loop gluon contribution ($ A B C = g t b $) dominates the standard
contribution in both the $g_{R}$ and $g_L$
dipole moments. All these diagrams, 19 as a whole, can be classified
according to their dependence on the quark masses.
\begin{figure}[ht]
\begin{center}
\includegraphics[width=8cm]{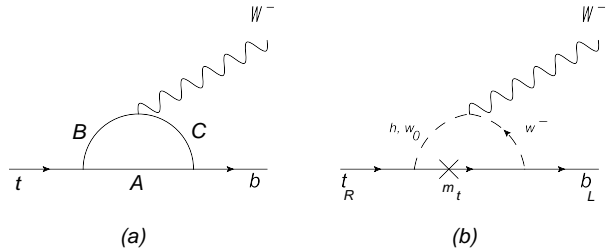}
\end{center}
\caption{
a. Topology of the one-loop SM Feynman diagrams for the quantum
correction to the decay $t\rightarrow b W^+$. 
b. Leading order diagrams for $g_R$ in the large $m_t$ limit.
}
\label{fig:higgs}
\end{figure}
As  already mentioned, the tensorial anomalous couplings we are interested
in are chirality
flipping magnitudes so, in general, a mass insertion is needed in order for the diagram
to contribute.  All contributions to $g_R$ need a $m_t$ mass insertion while
the contributions to $g_L$ need a $m_b$ mass insertion. The different mass
insertions for each diagram is shown in Figure 2. Besides, some of the
vertex have also a mass dependence. For all these
diagrams there are three mass dependencies that are different for the case
of  $g_R$ and for $g_L$.
We use this fact in order to classify all the contributions coming from all
the diagrams.
\begin{figure}[ht]
\begin{center}
\includegraphics[width=10cm]{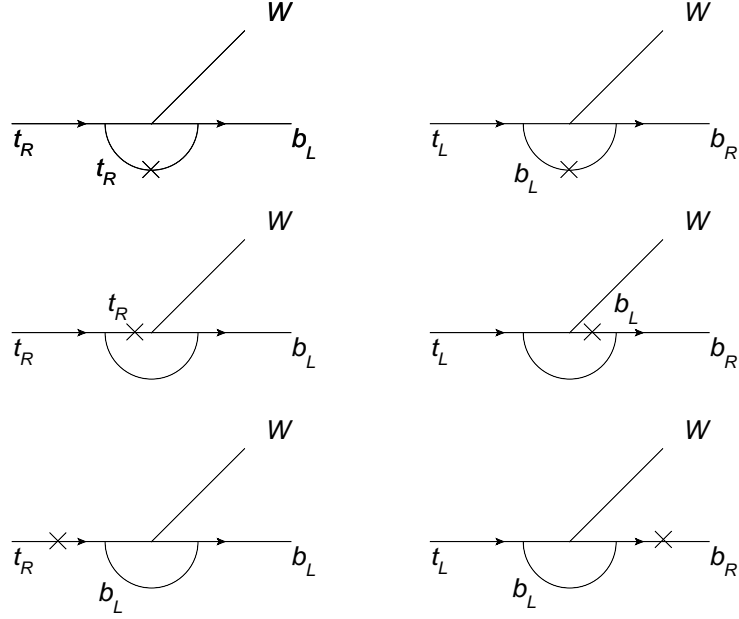}
\end{center}
\caption{
Mass insertions for the diagrams: in the left we can see the case
for $g_R$ where a factor $m_t$ is present for each diagram while the ones at
the right are for $g_L$ and proportional to  $m_b$.
  }
 \label{fig:higgs1}
\end{figure}
For $g_R$ there are two diagrams that
have a leading $m_t$-mass dependence, with non-decoupling in the $m_t$ mass
\cite{Bernabeu:1987me}. They are the ones with  $t h W$ and $t w_0 W$
circulating in the loop,
where $h$ is the Higgs boson  and $w_0$ is the
unphysical Z-boson; they are shown in Figure 1(b).
These two diagrams have top mass-insertions that, together with the mass
coming from the vertex, finally gives
a mass dependence that is of the order $1/r_w^2$, where
$r_w^2=(m_W/m_t)^2$,
with respect to the other diagrams.
Next, there are 12 diagrams  that also have a top mass insertion but do not
have this $1/r_w^2$
enhancement factor;   all of them have a similar mass dependence. The QCD
gluon exchange diagram also needs a top mass insertion for $g_R$.
The remaining  four diagrams do not have the $1/r_w^2$ enhancement factor
but a suppression factor  $r_b^2=(m_b/m_t)^2$ coming from the mass
dependence of the  vertex and from a b-quark mass insertion. The particles circulating in the
loop for these diagrams are $b w^+ w_0$,
$b w^+ h$, $h t b$ and $w_0 t b$, where $w^\pm$ are the unphysical W-bosons.
All these
last diagrams turn out to be numerically insignificant as the small $r_b$
coefficient may suggest.
Finally, the diagrams that dominate the final value for this electroweak
correction are the ones of the first two classes we have already presented.
In the
computation we find that there is no numerical domination of the first ones
with  the $m_t$ non-decoupling effect over the ones without this effect. The
$g_R$ coupling is finite and, up to one loop, the calculation needs no
renormalization. However, not all the diagrams are infrared finite: some
cancellation occurs among some groups of them in such a way as to end up with
a finite value. This fact was used as a check of our results. This is the
case for the diagrams $t \gamma W$ and $t W \gamma$ and also for $b \gamma
W$ and $b W \gamma$. There are no  singularities  when we sum each pair
together for the computation. Some diagrams, like $b W Z$ for example,
contribute to the imaginary part of $g_R$. In these cases we also used the
Cutkosky rules and compared the result with the direct computation as a
double check of our numerical calculation. Besides, some of the integrals
can be done analytically up to the end and we verified the numerical
evaluation of these results with the numerical evaluation of the Feynman
integrals. This check was possible  for the diagrams with a gluon or a
photon circulating in the loop. All these facts allow for a multiple check
of our calculations. As already anticipated, we can read in Table
\ref{bounds} that the value of the four diagrams with the $r_b^2$ factor
are numerical negligible.
\begin{table}[hbt]{\centering
\caption{Electroweak contributions to $g_R$ and $g_L$.}
\begin{center}
\begin{tabular}{||c||c|c|}
\hline
Diagram &$g_R$ $\times 10^3$ & $g_L$ $\times  10^3$\\ \hline
$t Z W$  &$-1.176$& $-0.0141$\\ \hline
$t h W$  &  $0.220$ & $0$ \\ \hline
$t  w^0 w^-$  &  $0.344$ & $0.0051$ \\ \hline
$t h w^-$  &  $0.462$ & $-0.0088$\\ \hline
$t Z w^-$& $-0.050$ & $-0.0012$ \\ \hline
$t \gamma W$ $+$ $t \gamma w^-$ & $0.572$ & $-0.0094$ \\ \hline
$b W Z$ & $- 0.623 - 0.664 i$ & $-0.0201 - 0.0214 i$  \\ \hline
$b W h$ & $0$ & $0.0086 - 0.0120 i$  \\ \hline
$b w^+ w^0$  & $(1.5+11.0 i)\times 10^{-4}$ & $-0.0029 - 0.0167 i$\\ \hline
$b w^+ h$ & $(-4.3+8.6 i)\times 10^{-4}$ & $-0.0019 + 0.0111 i$ \\ \hline
$b w^+ Z$ & $ -0.088 - 0.062 i$ & $-0.00039 - 0.00028 i$ \\ \hline
$b W \gamma$ $+$ $b w^+ \gamma$ & $0.114-0.509 i$& $-0.0270+0.0250 i$ \\ \hline
$Z t b $ & $-0.397$ & $-0.0067 $ \\ \hline
$\gamma t b$ & $0.068$ & $0.0115$ \\ \hline
$w^0 t b$ & $-6.8\times 10^{-4}$ & $-0.0109$ \\ \hline
$h t b $ & $-6.2\times 10^{-4}$ & $-0.0135$ \\ \hline \hline
$\Sigma (EW) $ & $- 0.56 - 1.23 i$ & $-(0.092+0.014 i)$\\ \hline
g t b & $-6.61$ & $-1.12$\\\hline \hline
Total& $-7.17 - 1.23 i$ & $-1.212 - 0.014 i$\\ \hline
\end{tabular}
\end{center}}
\label{results}
\end{table}
Each diagram contributes with a different sign to the final result so
finally many of them are numerically responsible for the final numerical
value that can not be anticipating without an explicit and careful
computation of all of them. The figures for each  contribution of the
diagrams to $g_R$ and $g_L$ is given in Table \ref{bounds}, where we take
the Higgs mass value $m_h = 150$ GeV. We can read in the Appendix A the
expressions for each of the diagrams as well as the analytical results (Appendix B) for
some of the diagrams for which it was feasible. We also show, in Appendix B, the limit $m_b
\rightarrow 0$ for these last formulas. This was also used as a check of
our numerical evaluation of the integrals. The results are rather
insensitive to the Higgs mass value in the experimental allowed Higgs mass
interval \cite{pdg}.
The final result for the one loop electroweak correction for the magnetic
right anomalous coupling is:
\beq
g_R^{EW} = - ( 0.56 + 1.23\, i ) \times 10^{-3} .
\label{eq:grew}
\eeq
Note that we have real and imaginary parts in this dipole moment and that
the last is more than double of the first. These values are to be compared
with the gluon contribution that is the dominant one:
\beq
g_R^g = - 6.61 \times 10^{-3} .
\label{eq:grg}
\eeq
This last result agrees with the one given in reference \cite{li} if we put
the numerical values for masses and couplings
used at that time.
The final result for the one loop computation in the SM is the sum
of the last two values given in Eqs.[\ref{eq:grew}] and
[\ref{eq:grg}]:
\beq
g_R^{SM} = - ( 7.17 + 1.23 i ) \times 10^{-3} .
\eeq
The real part for the one loop electroweak quantum correction for $g_R$ is
8\% of the leading gluon-exchange contribution. The CP-even imaginary absorptive part, generated by electroweak corrections, may be measured with a similar set of observables as those considered in the literature to measure $g_R$ and, more specifically, $Re(g_R)$. Note that this imaginary part is $17\%$ of the one loop $Re ( g_R^{SM})$.

The one loop  electroweak correction for the $g_L$  coupling can be obtained
in a similar way as in the previous calculation, however a b-quark mass
insertion is present in all the diagrams because $g_L$ couples to a right
b-quark.  This factor dominates the numerical value of the final result for
the $g_L$ electroweak contributions. As in the previous computation, there
are also IR divergences in the same diagrams as in the preceding calculation
and again  they sum up to a finite result. The same checks we already
explained before for $g_R$ have been used. We also find that an imaginary
part for the electroweak contribution to $g_L$  shows up, so the final
result is:

\beq
g_L^{EW} = - ( 0.92 + 0.14 i ) \times 10^{-4} .
\eeq
This is to be compared  and summed with the gluon contribution, that is real
and the dominant one, in order to obtain the one loop result:

\beq
g_L^g = - 1.12 \times 10^{-3} .
\eeq
The final result for the one loop computation in the SM is then:

\beq
g_L^{SM} = - ( 1.21 + 0.01 i ) \times 10^{-3} .
\eeq
We note that for $g_L$ the electroweak contribution is again 8\% of the
gluon contribution for $g_L$, and the CP-even imaginary part has its origin in the
electroweak diagrams.

\section{Conclusions}
\label{sec-conc}

We have computed the one loop electroweak  values of the
anomalous form factors $g_R$ and $g_L$ for the decay  $t \rightarrow b W^+$.
Both of them have real and imaginary parts. The imaginary parts come from
the electroweak correction and, for $g_R$, it is almost three times the real part, while for $g_L$, it is 15\% of the real one.
Contrary to what happens in extended models, where the imaginary part are
usually negligible if the new particles involved have higher mass scale than
the top, we found that the
absorptive parts of the dipole moments, which are induced in the SM by
$CP$-invariant final-state re-scattering, has to be taken into account and
may have physical effects that could be detected in the future through the observables proposed
in the literature. Note that the $g_R$ coupling will be measured at the LHC
and its absorptive contribution may be accessible in data and in the new
observables defined in \cite{pepe-ag}, for example. For the SM one loop $g_R$ coupling, the imaginary part 
is about 17$\%$ of the real one while, for $g_L$, it is only $1\%$. The value of the $g_L$ dipole moment, although proportional to $m_b$, is only about one order of magnitude smaller than $g_R$. The SM prediction
for the real part of $g_L$ is $Re ( g_L ) \simeq -0.0012$  and is very close
to the estimated 3$\sigma$ indirect discovery limit, $ Re ( g_L ) \leq
-0.0019$, obtained from $b\rightarrow s \gamma\ $ in \cite{pepe-ag} based on the results of \cite{Grzadkowski:2008mf}, so any contribution coming from new
physics that may show up may be in conflict with these bounds. Besides, the
value of the electroweak corrections for these dipole moments, first published in this paper can, by
themselves, explain deviations up to a few percent level in the observables, 
that are frequently discussed in the literature in connection to extended
models.

\subsubsection*{Acknowledgments}
This work was supported in part by Pedeciba and CSIC  - Uruguay, by
COLCIENCIAS - Colombia, by the Spanish
Ministry of Science and Innovation (MICINN), under grants FPA2008-03373,
FPA2008-02878, and by Generalitat Valenciana under grant 
PROMETEO 2009/128.
G.A.G.S. also thanks the Universitat de Valencia for the hospitality and
support during his stay at the Theoretical Physics Department.

\appendix
\section{Expressions for the diagrams}
For the $g_R$ and $g_L$ couplings the contribution coming from each of the
diagrams are written with the help of the following denominators:
\bea
&&\hspace*{-0.75cm} A_Z= x^2 \left(\left((y-1)  r_b^2+1\right) y-  r_w^2 (y-1)\right)-r_z^2 (x-1)\nonumber \\ 
&&\hspace*{-0.75cm} B_Z= x \left(\left((x (y-1)+1)  r_b^2+x-1\right) y-
r_z^2 (y-1)\right)- r_w^2 (x-1) (x (y-1)+1)\nonumber \\ 
&&\hspace*{-0.75cm} C_Z= (x-1) (x y-1)  r_b^2-r_w^2 (x-1) x (y-1)+ r_z^2 x
y+x (y-1) (x y-1)\nonumber \\ 
&& \hspace*{-0.75cm}\left\{A_\gamma,B_\gamma,C_\gamma\right\}
=\left\{A_Z,B_Z,C_Z\right\}(r_z\rightarrow 0)\nonumber \\ 
&& \hspace*{-0.75cm}\left\{A_H,B_H,C_H\right\}
=\left\{A_Z,B_Z,C_Z\right\}(r_z\rightarrow r_h)\nonumber
\eea

The contribution of each diagram to $g_L$ is:
\bea
&&\hspace*{-1cm}g_L^{tZW}=
\frac{e^2  V_{tb}^*\,  r_w\,  r_b }{128  \pi^2  s_w^2}\times
\int_0^1dx \int_0^1dy\;
\frac{
-2  ( a_t+ v_t) x \left(2 (y-1) y x^2-2 y x+x+1\right)}
{A_Z}\nonumber \\ 
&&\hspace*{-1cm}g_L^{t \gamma W}=
\frac{e^2  Q_t  V_{tb}^*\,  r_w\,r_b}{32  \pi^2   }\times
\int_0^1dx \int_0^1dy\;
\frac{
-2  x \left(2 (y-1) y x^2-2 y x+x+1\right)}
{A_\gamma}\nonumber \\ 
&&\hspace*{-1cm}g_L^{t h W} = 0\nonumber \\ 
&&\hspace*{-1cm}g_L^{t w_0 w^-}=
-\frac{e^2  V_{tb}^*\,  r_b}{128  \pi^2   r_w  s_w^2}\times
\int_0^1dx \int_0^1dy\;
\frac{
-2  x^3 y^2}
{A_Z}\nonumber \\ 
&&\hspace*{-1cm}g_L^{t h w^-}=
-\frac{e^2  V_{tb}^*\,  r_b}{128  \pi^2   r_w   s_w^2}\times
\int_0^1dx \int_0^1dy\;
\frac{
2  x^2 (x (y-2)+2) y}
{A_H}\nonumber \\ 
&&\hspace*{-1cm}g_L^{t Z w^-}=
\frac{e^2  V_{tb}^*\,  r_w\, r_b}{128   c_w^2 \pi^2    }\times
\int_0^1dx \int_0^1dy\;
\frac{
2 ( a_t+ v_t) (x-1) x}
{A_Z}\nonumber \\ 
&&\hspace*{-1cm}g_L^{t \gamma w^-}=
-\frac{e^2  Q_t  V_{tb}^*\,  r_w\, r_b}{32  \pi^2   }\times
\int_0^1dx \int_0^1dy\;
\frac{
2 x(x-1)}
{A_\gamma}
\nonumber \\ 
&&\hspace*{-1cm}g_L^{b W Z}=
\frac{e^2  V_{tb}^*\,  r_w\, r_b}{128  \pi^2   s_w}\times
\int_0^1dx \int_0^1dy\; \frac{2 x }{B_Z}\Big[
 a_b \left(2 (y-1) y x^2-5 y x+x+4\right)+\nonumber\\
&&\hspace{4cm} v_b \left(2 (y-1) y x^2+(y+1) x-2\right)\Big]
\nonumber \\ 
&&\hspace*{-1cm}g_L^{b W \gamma}=
\frac{e^2  Q_b  V_{tb}^*\,  r_w\, r_b}{32  \pi^2   }\times
\int_0^1dx \int_0^1dy\;
\frac{
2 x \left(2 (y-1) y x^2+(y+1) x-2\right)}
{B_\gamma}\nonumber \\ 
\nonumber \\ 
&&\hspace*{-1cm}g_L^{b W h}=
\frac{e^2 V_{tb}^*\,  r_w\, r_b}{64  \pi^2   s_w^2}\times
\int_0^1dx \int_0^1dy\;
\frac{
-2 (x-1) x}
{B_H}\nonumber\\
&&\hspace*{-1cm}g_L^{b w^+ w_0}=
-\frac{e^2 V_{tb}^*\,  r_b}{128  \pi^2  r_w  s_w^2}\times
\int_0^1dx \int_0^1dy\; \frac{
2 x^2 \left((x (y-1)+1)   r_b^2+x-1\right) y}
{B_Z}\nonumber \\
&&\hspace*{-1cm}g_L^{b w^+ h}=
-\frac{e^2 V_{tb}^*\,  r_b}{128  \pi^2  r_w  s_w^2}\times
\int_0^1dx \int_0^1dy\;
\frac{
2 x^2 \left((x (y-1)-1)   r_b^2-x+1\right) y}
{B_H}\nonumber \\
&&\hspace*{-1cm}g_L^{b w^+ Z}=
\frac{e^2  V_{tb}^*\,  c_w r_w\, r_b}{128  c_w^2 \pi^2}\times
\int_0^1dx \int_0^1dy\;
\frac{
-2 (  a_b-  v_b) x^2 (y-1)}
{B_Z}\nonumber \\ 
&&\hspace*{-1cm}g_L^{b w^+ \gamma}=
-\frac{e^2  Q_b  V_{tb}^*\,  r_w\, r_b}{32  \pi^2  }\times
\int_0^1dx \int_0^1dy\;
\frac{
2 x^2 (y-1)}
{B_\gamma}
\nonumber \\ 
&&\hspace*{-1cm}g_L^{Z t b}=
-\frac{e^2  V_{tb}^*\,  r_w\, r_b}{512  \pi^2  c_w^2  s_w^2}\times
\int_0^1dx \int_0^1dy\;
\frac{-4 ( a_t+ v_t) x^2 y}{C_Z}\left[ v_b (x-1)+ a_b (x+1)\right]
\nonumber \\
&&\hspace*{-1cm}g_L^{\gamma t b}=
-\frac{e^2  Q_b  Q_t  V_{tb}^*\,  r_w\, r_b}{32  \pi^2  }\times
\int_0^1dx \int_0^1dy\;
\frac{
-4 (x-1) x^2 y}
{C_\gamma}\nonumber \\ 
&&\hspace*{-1cm}g_L^{w_0 t b}=
\frac{e^2 V_{tb}^*\,  r_b}{128  \pi^2  r_w  s_w^2}\times
\int_0^1dx \int_0^1dy\;
\frac{
-2 x^2 (y-1) (x y-1)}
{C_Z}\nonumber \\ 
&&\hspace*{-1cm}g_L^{h t b}=
\frac{e^2  V_{tb}^*\,  r_b}{128  \pi^2  r_w  s_w^3}\times
\int_0^1dx \int_0^1dy\;
\frac{
2 x^2 (y-1) (x y+1)}
{C_H}\nonumber
\eea
while for $g_R$ we have:
\bea
&&\hspace*{-1cm}g_R^{t Z W}=
\frac{e^2  V_{tb}^*\,  r_w}{128  \pi^2  s_w^2} \times\nonumber \\ 
&&\displaystyle \int_0^1dx \int_0^1dy\; \frac{
2 x \left(v_t \left(-2 y x^2+x+1\right)+a_t \left(-2 y x^2+6 y
x+x-5\right)\right)}
{A_Z}\nonumber \\ [.5\baselineskip]
&&\hspace*{-1cm}g_R^{t \gamma W}=
\frac{e^2  Q_t  V_{tb}^*\,  r_w}{32  \pi^2   }\times
\int_0^1dx \int_0^1dy\; \frac{2 x \left(-2 y x^2+x+1\right)}
{A_\gamma}
\nonumber \\
&&\hspace*{-1cm}g_R^{t h W}=
\frac{e^2  V_{tb}^*\,   r_w}{64  \pi^2  s_w^2}\times
\int_0^1dx \int_0^1dy\;
\frac{
2 x^2 (1-y)}
{A_H}\nonumber \\ 
&&\hspace*{-1cm}g_R^{t w_0 w^-}=
-\frac{e^2  V_{tb}^*}{128  \pi^2   r_w  s_w^2}\times
\int_0^1dx \int_0^1dy\; \frac{
2 x^3 \left(-(y-1)  r_b^2-1\right) y}
{A_Z}\nonumber \\ 
&&\hspace*{-1cm}g_R^{t h w^-}=
-\frac{e^2  V_{tb}^*}{128  \pi^2   r_w   s_w^2}\times
\int_0^1dx \int_0^1dy\; \frac{
2 x^2 \left(x \left(1- r_b^2 (y-1)\right)-2\right) y}
{A_H}\nonumber \\ 
&&\hspace*{-1cm}g_R^{t Z w^-}=
\frac{e^2  V_{tb}^*\,   r_w}{128   c_w^2 \pi^2    }\times
\int_0^1dx \int_0^1dy\; \frac{
2 (a_t- v_t) (x-1) x}
{A_Z}\nonumber \\ 
&&\hspace*{-1cm}g_R^{t \gamma w^-}=
-\frac{e^2  Q_t  V_{tb}^*\,   r_w}{32  \pi^2   }\times
\int_0^1dx \int_0^1dy\; \frac{
-2x (x-1)}{A_\gamma}\quad(\mbox{diverge})\nonumber \\ 
&&\hspace*{-1cm}g_R^{b W Z}=
\frac{e^2  V_{tb}^*\,   r_w}{128  \pi^2   s_w}\times
\int_0^1dx \int_0^1dy\; \frac{
2 (a_b+ v_b) x \left(2 y x^2-(3 y+1) x+2\right)}
{B_Z}\nonumber \\ 
&&\hspace*{-1cm}g_R^{b W \gamma}=
\frac{e^2  Q_b  V_{tb}^*\,   r_w}{32  \pi^2   }\times
\int_0^1dx \int_0^1dy\; \frac{
2 x \left(2 y x^2-(3 y+1) x+2\right)}{B_\gamma}
\nonumber \\ 
&&\hspace*{-1cm}g_R^{b W h}=
\frac{e^2  r_b  V_{tb}^*\,   r_w}{64  \pi^2   s_w^2}\times 0 = 0
\nonumber \\ 
&&\hspace*{-1cm}g_R^{b w^+ w_0}=
-\frac{e^2  r_b  V_{tb}^*}{128  \pi^2  r_w  s_w^2}\times
\int_0^1dx \int_0^1dy\; \frac{
2 r_b x^3 y^2}
{B_Z}\nonumber \\ 
&&\hspace*{-1cm}g_R^{b w^+ h}=
-\frac{e^2  r_b  V_{tb}^*\,   r_w}{128  \pi^2  r_w^2  s_w^2}\times
\int_0^1dx \int_0^1dy\; \frac{
-2  r_b x^3 (y-2) y}
{B_H}\nonumber \\ 
&&\hspace*{-1cm}g_R^{b w^+ Z}=
\frac{e^2  V_{tb}^*\, s_w\, r_w}{128  c_w^2 \pi^2 }\times
\int_0^1dx \int_0^1dy\;
\frac{
-2 (a_b+ v_b) x^2 (y-1)}
{B_Z}\nonumber \\ 
&&\hspace*{-1cm}g_R^{b w^+ \gamma}=
-\frac{e^2  Q_b  V_{tb}^*\,  r_w}{32  \pi^2  }\times
\int_0^1dx \int_0^1dy\; \frac{
-2 x^2 (y-1)}{B_\gamma}
\nonumber \\ 
&&\hspace*{-1cm}g_R^{Z t b}=
-\frac{e^2  V_{tb}^*\,   r_w}{512   c_w^2 \pi^2 s_w^2}\times\nonumber \\ 
&&\int_0^1dx \int_0^1dy\; \frac{
-4 ( a_b+ v_b) x^2 (a_t (x (y-1)+2)+ v_t x (y-1)) y}
{C_Z}\nonumber \\ 
&&\hspace*{-1cm}g_R^{\gamma  t b}=
-\frac{e^2  Q_b  Q_t  V_{tb}^*\,   r_w}{32  \pi^2  }\times
\int_0^1dx \int_0^1dy\; \frac{
-4 x^3 (y-1) y}
{C_\gamma}\nonumber \\ 
&&\hspace*{-1cm}g_R^{w_0 t b}=
\frac{e^2 V_{tb}^*\, r_b}{128  \pi^2  r_w s_w^2}\times
\int_0^1dx \int_0^1dy\; \frac{
-2 r_b (x-1) x (x y-1)}
{C_Z}\nonumber \\ 
&&\hspace*{-1cm}g_R^{h t b}=
\frac{e^2  V_{tb}^*\, r_b}{128  \pi^2  r_w  s_w^2 }\times
\int_0^1dx \int_0^1dy\; \frac{
2  r_b (x-1) x (x y+1)}
{C_H} \nonumber
\eea

\section{Some exact results}
The following integrals, corresponding to diagrams with a photon or gluon circulating in the loop,
$t \gamma W$, $t \gamma w^-$, $b W \gamma$,
$b w^+ \gamma$, $\gamma t b$ and $g t b$, can be done analytically. Using
the notation
\[
g_{L,R}^{ABC}= \frac{e^2 V_{tb}^*\,    r_w}{32 \pi^2}\, Q\times I_{L,R}^{ABC}
\]
we have:
\bea
I_R^{t \gamma W + t \gamma w^-}&=&
\frac{4}{\Delta}
\left[\Bigg(
\left(1+\frac{1-r_w^2-r_b^2+\Delta}{4r_b^2}\right)
\log\left(\frac{1-r_w^2+r_b^2+\Delta}{1-r_w^2-r_b^2+\Delta}\right)\Bigg)-
\Bigg(\Delta\rightarrow -\Delta\Bigg)
\right]\nonumber \\ 
I_L^{t \gamma W + t \gamma w^-}&=&
\frac{2}{r_b}
\Bigg[1-\Bigg(\frac{\left(1-r_w^2+r_b^2+\Delta\right)
\left(1-r_w^2+3r_b^2+\Delta\right)}{4r_b^2\; \Delta }\;\log
\left(\frac{1-r_w^2+r_b^2+\Delta}{1-r_w^2-r_b^2+\Delta}\right)
\Bigg)\nonumber\\
&&\hspace{7cm}  - \Bigg(\Delta\rightarrow -\Delta\Bigg)\Bigg]\nonumber \\ 
I_R^{ t W \gamma + t w^+ \gamma}&=&
-\frac{2i\pi}{\Delta}\left[2-3r_w^2+r_w^4+r_b^4+r_b^2(1-2r_w^2)\right]
-2+2(2-r_w^2+r_b^2)\log(2r_b^2)+\nonumber \\ 
&&\hspace{1.5cm}
\frac{4r_b^2}{\Delta}\left[\Bigg(\frac{1-r_w^2+3r_b^2+\Delta}{(1-r_w^2+r_b^2+\Delta)^2}\; \log(1-r_w^2-r_b^2+\Delta)\Bigg)-\Bigg(\Delta\rightarrow
-\Delta\Bigg)\right]\nonumber\nonumber \\ 
I_L^{t W \gamma + t w^+ \gamma}&=&
\frac{2r_b}{\Delta}\left[\Bigg(\frac{1-r_w^2+3r_b^2+\Delta}{1-r_w^2 +
r_b^2+\Delta}\; \log\left(1-r_w^2-r_b^2+\Delta\right)\Bigg)-\Bigg(\Delta\rightarrow
-\Delta\Bigg)\right]-\nonumber\\
&&\hspace{2cm} 2r_b\log(2r_b^2)+\frac{2 i\pi r_b}{\Delta}(3-r_w^2+r_b^2)\nonumber \\ 
I_R^{\gamma t b }&=&
\frac{2}{\Delta}\left[\Bigg(\frac{1-r_w^2+r_b^2+\Delta}{1+r_w^2-r_b^2-\Delta}\; \log\left(\frac{2}{1-r_w^2+r_b^2+\Delta}\right)\Bigg)-\Bigg(\Delta\rightarrow
-\Delta\Bigg)\right]\nonumber\\
I_L^{\gamma t b }&=&
\frac{4r_b}{\Delta}\left[\Bigg(\frac{1}{1+r_w^2-r_b^2+\Delta}\; \log\left(\frac{2}{1-r_w^2+r_b^2-\Delta}\right)\Bigg)-\Bigg(\Delta\rightarrow
-\Delta\Bigg)\right]\nonumber
\eea
with $\dsp \Delta=\sqrt{1-2(r_w^2+r_b^2)+(r_b^2-r_w^2)^2}$.
These expressions can be written, in the $m_b\rightarrow 0$ limit, as:
\bea
I_R^{t \gamma W + t \gamma
w^-}&\approx&\frac{2}{1-r_w^2}\left[1+\frac{(2-r_w^2)}{1-r_w^2}\log(r_w^2)\right]+\nonumber\\
&&\frac{r_b^2}{(1-r_w^2)^4}\left[3(r_w^4-4r_w^2+3)+2(2+2r_w^2-r_w^4)\log(r_w^2)\right]+\mathcal{O}(r_b^4)\nonumber \\ 
I_L^{t \gamma W + t \gamma
w^-}&\approx&\frac{r_b}{(1-r_w^2)^3}\left[8r_w^2-3r_w^4-5+2(r_w^2-2)\log(r_w^2)\right]+\nonumber\\
&&
\frac{4r_b^3}{3(1-r_w^2)^5}\left[9r_w^2-r_w^6-8+3(r_w^4-2r_w^2-1)\log(r_w^2)\right]+\mathcal{O}(r_b^5)\nonumber \\ 
I_R^{t W \gamma + t w^+
\gamma}&\approx&-2i\pi\left[2-r_w^2+\frac{r_b^2}{(1-r_w^2)^2}(3-2r_w^2+r_w^4)\right]-2\left[1+(2-r_w^2)\log\frac{r_w^2}{1-r_w^2}\right]-\nonumber\\
&&\frac{2r_b^2}{(1-r_w^2)^2}\left[\log(r_b^2)-(3-2r_w^2+r_w^4)\log(1-r_w^2)+(2-2r_w^2+r_w^4)\log(r_w^2)\right]+\nonumber\\
&&\mathcal{O}(r_b^4)\nonumber \\ 
I_L^{t W \gamma + t w^+
\gamma}&\approx&2i\pi\frac{r_b}{1-r_w^2}\left(3-r_w^2+\frac{4r_b^2}{(1-r_w^2)^2}\right)+\nonumber\\
&&\frac{2r_b}{1-r_w^2}\left[\log(r_b^2)+(2-r_w^2)\log(r_w^2)-(3-r_w^2)\log(1-r_w^2)\right]-\nonumber\\
&&\frac{2r_b^3}{(1-r_w^2)^3}\left[1+2\log(r_b^2)+2\log(r_w^2)-4\log(1-r_w^2)\right]+\mathcal{O}(r_b^5)\nonumber \\ 
I_R^{\gamma t b
}&\approx&-\frac{2}{r_w^2}\log(1-r_w^2)+\frac{2r_b^2}{(1-r_w^2)^2}\left[1+\log(r_b^2)-2\log(1-r_w^2)\right]+\mathcal{O}(r_b^4)\nonumber \\ 
I_L^{\gamma t b
}&\approx&\frac{-2r_b}{1-r_w^2}\left[\log(r_b^2)-\frac{1+r_w^2}{r_w^2}\log(1-r_w^2)\right]-\nonumber\\
&&\frac{2r_b^3}{(1-r_w^2)^3}\left[1+2\log(r_b^2)-4\log(1-r_w^2)\right]+\mathcal{O}(r_b^5)\nonumber
\eea

\end{document}